\documentclass[12pt]{article}
\usepackage[english]{babel}
\usepackage{times}
\usepackage[T1]{fontenc}
\usepackage{epsfig}
\usepackage{amsmath}
\usepackage{amssymb}
\begin{document}

\title{Strictly finite range forces from the signum-Gordon field: exact results in two spatial dimensions}

\author{H. Arod\'z$^{\:a}$, J. Karkowski$^{\:a}$ and Z.
\'Swierczy\'nski$^{\:b}$ \\$\;\;$ \\ \emph{\small $^a$
Institute of Physics,
Jagiellonian University, Cracow, Poland }\\ \emph{\small
$^b$Institute of Computer Science and Computer Methods,}\\\emph{\small Pedagogical University, Cracow, Poland}}

\date{$\;$}

\maketitle

\begin{abstract}
Exact formula for the force between two identical static  point
charges coupled to  the  nonlinear scalar field of  two-dimensional signum-Gordon
model is obtained.  Pertinent solution of  the field equation is found in the form of one dimensional integral.   The force exactly vanishes when the distance between charges exceeds certain critical value.   
\end{abstract}

\vspace*{2cm} \noindent PACS: 11.27.+d, 11.10.Lm, 03.50.Kk
\\

\pagebreak

\section{ Introduction}

Nonlinearity of field equations can have  pronounced manifestations in physical predictions from the theory. Examples are ubiquitous and well-known:  multiple ground states, static solitons, solitary waves, long lived oscillons,  blow-up of solutions at a finite time, and many more.  Another important  effect due to the  nonlinearity is  that the field can react  to  the presence of external charges in a rather nontrivial way. In consequence,  the force between the charges essentially differs from  naive expectations based on free field models. Very old, yet still interesting example of this phenomenon  is provided by classical  non-Abelian gauge fields in the presence of  static external charges, see, e.g.,
\cite{1},  \cite{2}, \cite{3}.  Recently, we have studied such this aspect of the nonlinearity using  the signum-Gordon model, \cite{4}, \cite{5}.  The remarkable simplicity of the field equation in this model enables us to present exact nonperturbative results, in contradistinction to the case of Yang-Mills field.  The most striking findings are as follows:  there are no  
Yukawa or Coulomb tails of the field in the region far away from charges;  the charges are totally screened by the field;  the force between  them exactly vanishes  when they are separated by a distance that exceeds a certain finite value.  In the present paper we corroborate these results by solving the two dimensional case.    

 Let us briefly remind that the signum-Gordon model involves a single real 
scalar field $\varphi$ that evolves according to the signum-Gordon equation
\[\partial_{\mu} \partial^{\mu} \varphi + g\: \mbox{sign}\: \varphi  = 0,\]  where $g>0$ is the self-coupling constant.  The $\mbox{sign}$ function takes the values $0, \pm1$,  in particular $\mbox{sign}\; 0=0$.  The Lagrangian has the form
\[ L = \frac{1}{2} \partial_{\mu} \varphi \:
\partial^{\mu} \varphi
 - g \:|\varphi|.  \]
The field potential  $U(\varphi) = g |\varphi|$ is V-shaped  ($|\;|$  denotes the modulus).  The model has been studied in several aspects, such as self-similar solutions \cite{6}, oscillons (or rather breathers) \cite{7},  and $Q$-balls or boson stars in the version with a complex scalar field \cite{8}.  Let us add that recently the signum-Gordon equation has been discussed in a much wider mathematical framework of the theory of partial differential equations with compressed solutions \cite{9}. In general, the model together with its extensions has
turned out to be very useful theoretical laboratory in which we can study various above mentioned  aspects of nonlinear fields.

In the present paper,  similarly as in  \cite{4}, \cite{5}, we regard the signum-Gordon field as the mediating field, which generates a force between two point charges coupled to it.  In \cite{4}, the field generated by the two charges, as well as the force, are exactly calculated in one spatial dimension, while  in three dimensions approximate formulas are given under the assumption that the point charges are close to each other.  In \cite{5}  we study three-body forces in the one dimensional case.  Because the signum-Gordon equation is nonlinear one, finding its solutions is generally a nontrivial task, especially  in dimensions larger than one. In the present paper we investigate the two dimensional case. Using a formal mathematical connection with the planar electrostatics we show that also in this case the point charges can be totally screened by a cloud of the scalar field. This screening is easily seen for a single charge. The case of two or more charges situated not too far from each other is much more  difficult. The main problem is the determination of the shape of the screening cloud. We have found this shape, as well as  an integral formula for the scalar field  forming the cloud. These results are used in order to derive the exact formula for the force between the two identical charges.   The force completely vanishes when the distance between the charges exceeds certain critical value (equal to $2 R_0(q)$, see formula (4) below) that  non-analytically depends  on the strength $q$ of the charges and  the self-coupling constant  $g$ .

The plan of our paper is as follows. In Section 2 we establish the presence of the total screening in the case of  two identical charges in two spatial dimensions. We find the exact shape of the screening cloud formed by the scalar field, and we obtain the integral formula for the field. Section 3 is devoted to the calculation of the force between the charges. Section 4 contains a summary and remarks.

\section{The total screening of charges  }

Let us begin from the simple case of a  single point-like charge of strength $q>0$
located at the origin in the two dimensional space.  The field equation for the static field has the form
\begin{equation}
\triangle \varphi  =  g\: \mbox{sign}\: \varphi  - q \delta(\vec{x}),
\end{equation}
where $\triangle$  is the two dimensional Laplacian. 
The fundamental solution of  the linear Poisson equation  $\triangle G = - q \delta(\vec{x})$ has the form
\[ G(\vec{x}) = - \frac{q}{2\pi} \ln\frac{|\vec{x}|}{l_0}, \]
where  $l_0$ is a constant.  We expect that close to the charge the solution $\varphi$ is  well approximated by $G(\vec{x})$, and therefore it is positive.  Then the  term $ g\: \mbox{sign}\: \varphi $ in (1)  has the constant value $g$, and the exact solution of Eq.\ (1) has the form
\begin{equation}
 \varphi(\vec{x})  = G(\vec{x}) + \frac{g}{4} \vec{x}^{\:2} +c_0.
\end{equation}
Here $c_0$ is another constant.   Because it can be included into the constant $l_0$,  we put $c_0=0$.    The constant  $l_0$ is determined from the  requirement that our $\varphi(\vec{x})$  matches the vacuum solution $\varphi=0$ at a certain  radius  $ R_0$.  The matching conditions have the standard form,
\[ \varphi=0, \;\;\;   \varphi' =0  \]
on the circle  $|\vec{x}|=R_0$.  Here $'$ stands for the derivative with respect to $|\vec{x}|$. 
The final form of  solution (2)  reads
\begin{equation}
 \varphi  = - \frac{q}{2\pi} \ln\frac{|\vec{x}|}{R_0(q)}  + \frac{g}{4} \vec{x}^{\:2} - \frac{q}{4\pi},
\end{equation}
where \begin{equation} R_0(q)= \sqrt{\frac{q}{\pi g}}.\end{equation}
 The field $\varphi$ given by formula (3) has  positive values
in the circle  $|\vec{x}| <R_0(q)$,  hence the assumption  that
$\mbox{sign}\varphi =+1$ is fulfilled.  To summarize, the scalar field in the presence of the single charge has the form 
\[ \varphi_q(\vec{x}) = \left\{ \begin{array}{ccc} - \frac{q}{2\pi} \ln\frac{|\vec{x}|}{R_0(q)}  + \frac{g}{4} \vec{x}^{\:2} - \frac{q}{4\pi}
 & \mbox{for}
&|\vec{x}| < R_0(q),  \\  0 &  \mbox{for} &|\vec{x}| \geq R_0(q). 
\end{array}         \right.\]

The field  $\varphi_q(\vec{x})$  is  quite interesting  for the following  reasons.  First, note that its dependence on the coupling constant
$g$, as well as on the charge $q$, is nonanalytic at $g=0,\; q=0$.  Thus  the solution is non perturbative in  these parameters.   Second,   the charge $q$ is completely
screened by the circular cloud of the field, because   outside the circle  $|\vec{x}| \leq R_0$  the
field attains the vacuum value $\varphi=0$  exactly. The cloud has the
constant charge density $g \:\mbox{sign}\varphi =g$ of the opposite
sign than $q$.  Formula (4)  for the radius $R_0(q)$  is equivalent to the statement 
 that the total absolute charge of the circular cloud is
equal to $q$, namely $\pi R_0^2(q) g=q$.

Now let us turn to  less trivial case of two identical point
charges of strengths $q$ located  at the points $ \pm \vec{b}= (\pm
b, 0)$ on the $x^1$ axis in the two dimensional space. The distance between charges is  given by
$d=2b$.   Instead of Eq.\ (1) we now have
\begin{equation}
\triangle \varphi  =  g\: \mbox{sign}\: \varphi  - q \delta(\vec{x}- \vec{b}) -q \delta(\vec{x}+ \vec{b}) .
\end{equation}
Because of the nonlinear $\mbox{sign}\:\varphi$ term, the  pertinent solution of this equation is not just the
sum of appropriately shifted in space solutions (3),  unless the
distance $d$ between the charges exceeds  $2R_0(q)$, in which case
the two circular screening clouds surrounding the charges do not intersect each other.  If the charges are close to each other, i.e.,
$d<2R_0(q)$,  it is not  clear whether the total screening is still present, and if present, what is the shape of the  screening cloud of
the field. We address these questions in the remaining part of  this  section.  Our findings  are  utilized in the next  section,
where we calculate the force exerted on the charge located at
$\vec{x}= \vec{b}$.

It is clear that when $b=0$ we again have a single point charge, screened by the circular cloud of the field
as discussed above, except that now the strength of the charge equals $2q$ instead $q$.  
 Let us
assume for a moment that the total screening persists also when
$b>0$, that is that there exists a region $\Sigma$ surrounding the two
charges  such that $\varphi(\vec{x})=0$ for points $\vec{x}$ lying
outside $\Sigma$, or on the boundary $\partial \Sigma$, and  that $\varphi(\vec{x})>0$ if $\vec{x}$ lies inside $\Sigma$.   Inside
$\Sigma$, the term $ g\: \mbox{sign}\: \varphi =g $ in Eq.\ (5)
provides the constant charge density that screens the two point
charges. This cloud of charge  contributes to $\varphi(\vec{x}) $
the term
\begin{equation}
\varphi_{cloud}(\vec{x}) = \frac{g}{2\pi} \int_{\Sigma} d^2y \: \ln\frac{|\vec{x}-\vec{y}|}{R_0(q)}.\end{equation}
The two point charges contribute
\begin{equation}\varphi_0(\vec{x}) = - \frac{q}{2\pi}\ln\frac{|\vec{x}-\vec{b}|}{R_0(q)} - \frac{q}{2\pi}\ln\frac{|\vec{x}+\vec{b}|}{R_0(q)}.\end{equation}
Note that $\varphi_0(\vec{x})$ is the solution of (5) when $g=0$. The shape of the region $\Sigma$ can be found  from the requirement that we have the total screening, i.e., 
\begin{equation}
\varphi_{cloud}(\vec{x}) + \varphi_0(\vec{x}) =0
\end{equation}
for all points $\vec{x}$ outside $\Sigma$.   

As already noticed, Eq.\ (5) is similar to Poisson equation of ordinary electrostatics. Standard reasoning known from  the electrostatics gives  the boundary conditions for $\varphi$ at $\partial \Sigma$:  $\varphi(\vec{x})=0, \; \partial_n \varphi(\vec{x}) =0$ for all $\vec{x} \in \partial \Sigma$,  where $\partial_n$ denotes the derivative in the direction perpendicular
 to $\partial \Sigma$.  These conditions are checked numerically after we determine  the region $\Sigma$.

The condition (8) is  utilized as follows.  First, we rewrite the formulas for $\varphi_{cloud}(\vec{x}), \: \varphi_0(\vec{x})$ using the rescaled polar coordinates, introduced as follows:
\[ \vec{x}= R_0(q) \:r\: \left( \begin{array}{c} \cos\theta\\ \sin\theta \end{array} \right), \;\; \;\;\vec{y}= R_0(q) \:\rho\: \left( \begin{array}{c} \cos\alpha\\ \sin\alpha \end{array} \right).    \]
In particular,  $r= |\vec{x}|/ R_0(q)$ and  $\rho= |\vec{y}|/ R_0(q)$. 
Thus, 
\[\varphi_{cloud}(r, \theta) = \frac{q}{4\pi^2}\: \int^{2\pi}_0\!\!\!\! d\alpha\:\int^{r_0(\alpha)}_0\! \!\!\!d\rho\: \rho \:[2 \ln r + \ln(1+\frac{\rho^2}{r^2} - \frac{2\rho}{r}\cos(\theta-\alpha))], \]
\[ \varphi_0(r, \theta)= - \frac{q}{\pi} \ln r - \frac{q}{4\pi} \ln(1+\frac{d^2}{r^2} - \frac{2d}{r}\cos\theta) - \frac{q}{4\pi} \ln(1+\frac{d^2}{r^2} + \frac{2d}{r}\cos\theta),  \]
where
\[ d=\frac{b}{R_0(q)}. \] The function $r_0(\alpha)$  gives the radial coordinate of the boundary of $\Sigma$ at the  azimuthal angle  $\alpha$.  The considered set  of two point charges is symmetric with respect to reflections in the both axises,  as well as in the origin. We expect that the shape of  $\Sigma$  reflects these symmetries. Therefore,
\[ r_0(\alpha) = r_0(2\pi -\alpha) =  r_0(\pi -\alpha)= r_0(\pi +\alpha). \]

Our present task is to determine the function  $r_0(\alpha)$. To this end,  we first notice that
\[ \ln(1+\frac{\rho^2}{r^2} - \frac{2\rho}{r}\cos(\theta-\alpha)) = \ln(1 - \frac{\rho}{r}\exp[i(\theta-\alpha)]) + \ln(1 - \frac{\rho}{r}\exp[-i(\theta-\alpha)]).\] Expanding the two logarithms on the r.h.s with respect to $\rho/r$,  integrating over $\rho$, and using the symmetry of $r_0(\alpha)$ we  obtain
\begin{equation}
\varphi_{cloud}(r, \theta) = \frac{q}{4\pi^2}\: \int^{2\pi}_0\!\!\!\! d\alpha\;r_0^2(\alpha) \:\ln r -  \frac{q}{2\pi^2}\sum^{\infty}_{k=1}\frac{\cos(k \theta)}{k (k+2) r^k} \int^{2\pi}_0\!\! \!\!d\alpha\;r_0^{k+2}(\alpha) e^{ik\alpha}.
\end{equation}
This formula is to be compared with the Fourier series in $\theta$ for  $ \varphi_0(r, \theta)$, which   can be obtained by expanding the logarithms similarly as above, and reads
\begin{equation}
 \varphi_0(r, \theta) = - \frac{q}{\pi} \ln r + \frac{q}{2\pi} \sum^{\infty}_{k=1}\frac{1+(-1)^k}{k} \frac{d^k}{r^k}\cos(k \theta).
\end{equation}
Condition (8)  is satisfied if
\begin{equation}
 \int^{2\pi}_0\! \!\!d\alpha\:r_0^{2}(\alpha) = 4 \pi,
\end{equation}
and
\begin{equation}
 \int^{2\pi}_0\! \!\!d\alpha\:r_0^{k+2}(\alpha)\: e^{ik\alpha}= (1+(-1)^k)\: (k+2)\: \pi\: d^k,
\end{equation}
where $k=1,2,\ldots$. Because $r_0(\alpha)= r_0(\pi-\alpha)$, conditions (12) are satisfied automatically for odd values of $k$. For even values of $k$,  we put $k=2l$ in (10)
 and rewrite (11) and  (12) as the following set of conditions for $r_0(\alpha)$
\begin{equation}
 \int^{2\pi}_0\! \!\!d\alpha\:r_0^{2(l+1)}(\alpha)\: e^{2il\alpha}= 4  (l+1)\: \pi\: d^{2l},
\end{equation}
where $l=0,1,2, \ldots$.

 Because the r.h.s.'s of  conditions (13) depend on $d^2$, we expect that $r_0^2(\alpha)$  contains only  even powers of $d$,
\[ r_0^2(\alpha) = f_0(\alpha) + d^2 f_2(\alpha) + d^4 f_4(\alpha) + \ldots. \]
The functions $f_{2n}(\alpha) $, $n=0,1,2, \ldots$, have Fourier representation of the general form
\begin{equation} f_{2n}(\alpha) = \sum^{\infty}_{k=0}c_{2n;k}\cos(2k\alpha),\end{equation}
in compliance with the symmetries of $r_0(\alpha)$.  The first function, $f_0$,  gives $r_0^2$ when $d^2=0$. Thus, we put $d^2=0$ and  $r_0^2(\alpha) = f_0(\alpha)$  in  conditions  (13); $f_0(\alpha)$ has the Fourier form  (14).  Simple analysis based on the formulas $2\cos(2k\alpha) = \exp(2ik\alpha) + \exp(-2ik\alpha) $ and $ \int^{2\pi}_0 d\alpha \exp(i n\alpha)  = 2\pi \delta_{n0}$    shows  that  the conditions for $f_0(\alpha)$ are satisfied only if  this function is constant,  $f_0(\alpha)=2$.   This result is in agreement with the fact that for $d=0$ we actually have the single point charge of the strength $2q$, which has the circular screening cloud of the radius $R_0^2(2q)= 2R_0^2(q)$, i.e., $r_0^2=2$ for our rescaled radial coordinate.

 In order to determine  the function $f_2(\alpha)$, we differentiate  conditions  (13) with respect to $d^2$, and we put $d^2=0$, $f_0=2$. This gives the conditions
\[  \int^{2\pi}_0\! \!\!d\alpha\:\: e^{2il\alpha}f_2(\alpha) = 2 \pi\:\delta_{l1},
\]where $l=0,1,2, \ldots.$ It follows that $f_2= 2 \cos(2\alpha)$.  Taking higher derivatives of (13) with respect to $d^2$ and performing  similar calculations  as above, we have found that $f_4=0, \; f_6=0, \; f_8=0. $ With such partial results, we  have made the educated guess that  $f_{2n}=0$ for all $n \geq 2$, i.e., that 
\begin{equation}
r_0^2(\alpha) = 2\: [1+d^2 \cos(2\alpha)].
\end{equation}
 It turns out that indeed,  $r_0^2(\alpha)$ given by this formula  satisfies all conditions  (13).  The pertinent integration on the l.h.s. of (13) is elementary.

The form of  formula (15) implies that it holds only if $d\leq1$, because  $r_0^2 \geq 0$.  In the case $d=1$,  i.e., $|\vec{b}|= R_0(q)$,  this formula 
 gives  two circles which touch each other at the origin, and have the point charges at their centers.  For $d>1$  each charge has its own circular
 screening cloud, separated  from the other.

It remains to check whether all values of the total field
$\varphi(\vec{x}) = \varphi_{cloud}(\vec{x}) +\varphi_0(\vec{x})$
are strictly positive   inside the region $\Sigma$, as it has been
assumed. We can do this only numerically. It is a rather
straightforward computation since the integral (6) giving
$\varphi_{cloud}$ has the already known compact domain $\Sigma$. Moreover,  the two dimensional integral can  be reduced to
one-dimensional one over the boundary of $\Sigma$, see formula (19)
below.  The numerical results  corroborate our assumption.  They also show that $\varphi$ and $\partial_n\varphi$ are continuous  on $\partial\Sigma$.

Thus, we conclude that the pertinent solution of Eq.\ (5),
denoted below as $\varphi_{qq}(\vec{x})$, has the following form
\begin{equation}
\varphi_{qq}(\vec{x}) = \left\{\begin{array}{ccc}
\varphi_{cloud}(\vec{x}) +\varphi_0(\vec{x}) & \mbox{for}
&\vec{x}\in \Sigma,  \\  0 &  \mbox{for} &\vec{x}\notin \Sigma,
\end{array} \right.
\end{equation}
where the compact region $\Sigma$   has the boundary with the rescaled radial coordinate $r_0(\alpha)$
given by formula (15). Example
plots of levels of $\varphi_{qq}(\vec{x})$ are presented in Figs. 1 and 2.

\begin{center}
\begin{figure}[tph!]
\hspace*{1.5cm}
\includegraphics[height=8cm, width=9.5cm]{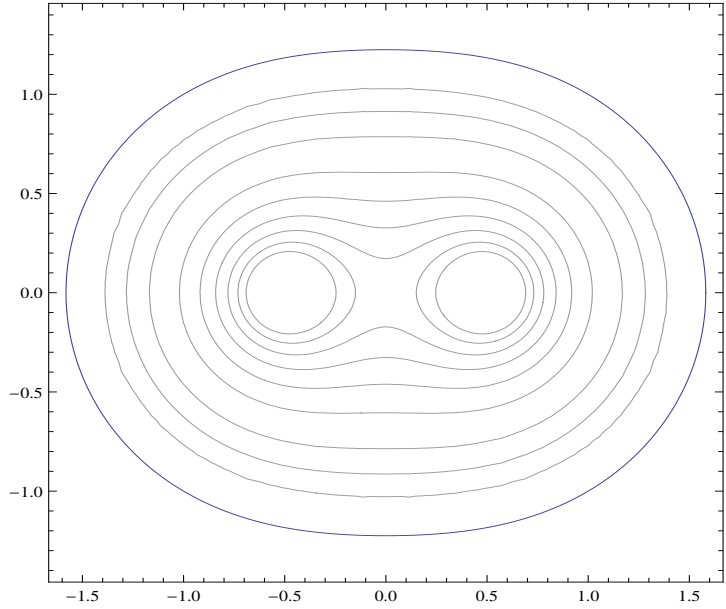}
\caption{\small The contour plot of  $\varphi_{qq}$ for $d=1/2$. The outermost  contour  corresponds to $\varphi=0$ and it is given by $r_0(\alpha)$,  formula (15).  This contour encircles the compact domain $\Sigma$. The horizontal and vertical axises correspond to $x^1/R_0(q)$, $x^2/R_0(q)$, respectively.   }
\end{figure}
\end{center}

\begin{center}
\begin{figure}[tph!]
\hspace*{1cm}
\includegraphics[height=6cm, width=11cm]{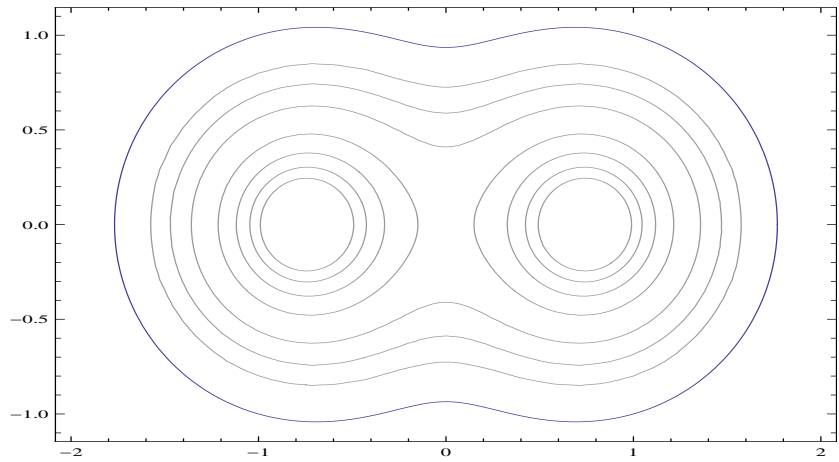}
\caption{\small The contour plot of  $\varphi_{qq}$ for $d=3/4$. The meaning of the lines is the same as in Fig. 1.   The picture shows the deformation of outer contours in vicinity of the vertical  line $x^1=0$, that ultimately leads to the breakup of the cloud into two non overlapping  circular clouds when $d=1$.    }
\end{figure}
\end{center}

\section{The force exerted on the charge located at $\vec{x}=\vec{b}$ }

The method of calculating the force exerted on an external charge
coupled to a field  is discussed in detail in  \cite{4}. Adapting it
to  the case at hand, we extract from the field $\varphi_{qq}$  the
logarithmically divergent  proper field of the charge located  at
$\vec{x}=\vec{b}$,
\[ \varphi_{qq}(\vec{x}) = - \frac{q}{2\pi} \ln\frac{|\vec{x}-\vec{b}|}{R_0(q)} + u(\vec{x}).\]
Here  \[u(\vec{x}) =  - \frac{q}{2\pi}
\ln\frac{|\vec{x}+\vec{b}|}{R_0(q)} + \varphi_{cloud}(\vec{x})\] is
smooth at the point $\vec{x}=\vec{b}$.  Next,  we calculate the time rate of the 
transfer of momentum from the field to the charge located at
$\vec{x}=\vec{b}$. It is equal to the force $\vec{F}$
exerted on that charge,  see \cite{4} for details and examples.   It turns out that
\begin{equation}
\vec{F} = \left. q \nabla u \right|_{\vec{x}=\vec{b}},
\end{equation}
where $\nabla u = ( \partial u/ \partial x^1,  \partial u/ \partial
x^2 )$. We see that  the proper field of the charge does not contribute to the
force exerted on that charge, as expected.  The symmetry of the set of charges implies that
$F^2=0$. The formula for the non vanishing component $F^1$ can be
rewritten in the polar coordinates as
\begin{equation}
F^1= q \left.  \frac{\partial u}{\partial
|\vec{x}|}\right|_{\stackrel{\theta=0}{ |\vec{x}|=b}} = \frac{
q}{R_0(q)}\: \left. \frac{\partial u}{\partial
r}\right|_{\stackrel{\theta=0}{r=d}}.
\end{equation}

Computation of the force from formula (18) is significantly
simplified when we rewrite $\varphi_{cloud}$ inside the region $\Sigma$  as a line integral
over the boundary $\partial \Sigma$. First, we  write  $ \varphi_{cloud}$ in the form 
\[ \varphi_{cloud}(\vec{x}) = \frac{g}{8\pi} \int_{\Sigma}\!\! d^2y\: (\partial_{y^1}W^2-\partial_{y^2} W^1), 
\]
where \[ W^i(\vec{x}, \vec{y}) = \epsilon^{ik} (x^k-y^k)  (\ln\frac{(\vec{x}-\vec{y})^2}{R_0^2(q)} - 1), \]  \[
  \partial_{y^1}W^2-\partial_{y^2} W^1 = 2 \ln\frac{(\vec{x}-\vec{y})^2}{R_0^2(q)}, \]
and $\epsilon^{ik}$ is the antisymmetric symbol,  $i,k=1,2$.   Next we apply Stokes  theorem, 
\begin{equation}
 \varphi_{cloud}(\vec{x}) =
  \frac{g}{8\pi}\oint_{\partial\Sigma}\!\!\! d\alpha   \;\partial_{\alpha} \vec{y}_0(\alpha)\; \vec{W}(\vec{x}, \vec{y}_0(\alpha)),
\end{equation}
The curve $\vec{y}_0(\alpha) $,  defined by
 \[ \vec{y}_0(\alpha) = R_0(q)\: r_0(\alpha)\:
\left(\begin{array}{c} \cos\alpha\\ \sin\alpha  \end{array} \right),   \]
represents $\partial \Sigma$.   Using the contour representation (19) we obtain
\begin{equation}\left. \frac{\partial \varphi_{cloud}}{\partial
r}\right|_{\stackrel{\theta=0}{r=d}}= \frac{g R_0^2(q)}{4\pi}
 \left[ \int^{2\pi}_{0}\!\! d\alpha\:  h_1(\alpha)\: h_2(\alpha) + \frac{1}{2} \int^{2\pi}_{0}\!\! d\alpha\:  h_3(\alpha)\: h_4(\alpha)\right],
 \end{equation}
where
\[
h_1(\alpha)=  \frac{d- r_0(\alpha) \:\cos\alpha }{d^2-2\:d\:
r_0(\alpha)  \:\cos\alpha + r_0^2(\alpha)}, \]
\[ h_2(\alpha)= r_0^2(\alpha) - d  \: r_0(\alpha)\: \cos\alpha  +\frac{2 d^3 \sin\alpha\: \sin(2\alpha)}{r_0(\alpha)},
\]
\[
h_3(\alpha)= \ln[d^2 - 2\:d\:r_0(\alpha)\: \cos\alpha +
r_0^2(\alpha)] -1,
\]
\[
h_4(\alpha)= \frac{2\:d^2 \sin\alpha \sin(2\alpha)}{r_0(\alpha)} -
 r_0(\alpha)\:\cos\alpha, \]  $r_0(\alpha)$ is given by formula
(15),  $0<d<1$.  

 In spite of their appearance, the integrals in (20) are elementary.
First, notice that $h_4(\alpha)= -
\partial_{\alpha}(r_0(\alpha)\:\sin\alpha)$. Therefore,  in the
second integral in (20) we may use integration by parts in order to
eliminate the logarithm function. Second, we write the $\sin$
and $\cos$  functions in terms of $\exp(\pm i \alpha)$ and observe
that $\int^{2\pi}_0 d\alpha \exp( i k \alpha)=0$ for any integer $k$
except $k=0$. This eliminates many terms, in particular all terms
proportional to $\sqrt{1+d^2 \cos(2\alpha)}$. This last function is thought of as the Taylor series with respect to $d^2 \cos(2 \alpha)$,   obviously it contains only even powers of $\exp(\pm i \alpha)$.  Finally, we use the
integral 
\[\int_0^{2\pi}\! \frac{d \alpha}{4+ d^4 +4d^2\:\cos(2\alpha)} =
\frac{2\pi}{4-d^4},
\]
and formula (4) for  $R_0(q)$. 
We obtain
\begin{equation}\left. \frac{\partial \varphi_{cloud}}{\partial
r}\right|_{\stackrel{\theta=0}{r=d}}= \frac{q}{4\pi} \:d\:(2-d^2).
\end{equation}

The non vanishing component of the force exerted on the charge located at the point $x^1=b, \: x^2=0$ is computed from formula  (18),
\begin{equation}
F^1= - \frac{q^2}{4\pi b} \left(1- \frac{b^2}{R_0^2(q)}\right)^2 = - \frac{q^2}{4\pi b} \left(1- \frac{\pi g b^2}{q}\right)^2.
\end{equation}
Here we have returned to the original  coordinate $b= R_0(q) d$. 
The force is attractive one, as expected from the scalar field. Formula (22) holds for $b <  R_0(q)$, outside this range of $b$ the force vanishes \footnote{$R_0(q)$ vanishes in the limit $q\rightarrow 0$. Therefore, the force also vanishes in this limit, in spite of the fact  that the r.h.s. of formula (22)  gives  $-\pi g^2 b^3/4$.  The point is that in that limit  formula (22)  is not valid for any $b $  -- the force vanishes for all $b \geq R_0(q)$.  }. 

The factor  $-q^2/4\pi b$ in formula (22) represents the standard two dimensional Coulomb force  characteristic for the free field  ($g=0$).  This force dominates at short distances,  $l\ll l_0$,  also in the case of self-interacting field. Here $l=2b$ is the distance between the charges, and $l_0=2 R_0(q)$ is the critical distance between them.   At  distances $ l \lessapprox  l_0    $  the force becomes weaker, and  it exactly vanishes  for $l \geq l_0$.  When the distance $l$ reaches $l_0$, the screening cloud splits into two non-overlapping  circular clouds that screen each charge separately. 

Formula  (22)  for the force can easily be generalized to the case one particle is located at the point $\vec{x}$ and the other one at $\vec{y}$.  It is sufficient to substitute $b= l /2$, where  $l=|\vec{x} - \vec{y}\:|$,   and to include the  unit  vector $\vec{n} = (\vec{x} - \vec{y}\:)/  |\vec{x} - \vec{y}|$  directed from $\vec{y}$ to $\vec{x}$. Thus, the force exerted on the particle located at $\vec{x}$ is given by formula
\begin{equation}
\vec{F}(\vec{x}) =  - \frac{q^2}{2\pi l}\left(1- \frac{  l^2}{l_0^2}\right)^2 \:\vec{n}
\end{equation}
if $l=|\vec{x}-\vec{y}|  <  l_0$,  otherwise  $\vec{F}(\vec{x}) =0$.  
This force  possesses the potential  $U_{qq}$  such that  $\vec{F}(\vec{x}\:) = - \partial_{\vec{x}} U_{qq}$, namely 
\[  U_{qq} =  \frac{q^2}{2\pi } \left[\ln \frac{l}{l_0} - \left(\frac{l}{l_0}  \right)^2   +\frac{1}{4} \left(\frac{l}{l_0}  \right)^4 + \frac{3}{4}   \right]   \]  if $l < l_0$,  otherwise $U_{qq}=0$.

\section{Summary and remarks}

1.    As the most  valuable result of our work we regard the exact formula (23) for the force  exerted by one point charge on the other.   This force  vanishes quadratically when  the distance $l$  between the charges approaches (from below) the critical value $l_0= 2 \sqrt{q/\pi g}$. In the one dimensional case investigated in \cite{4}, the force vanishes linearly, $F= q^2\: (1-a/a_{*})/2$, where $a$ is the distance between the charges, and the critical distance $a_{*}$ is given by the formula $a_{*}= q/g$ (we keep the same form of  the field equation in all dimensions).  The situation in three dimensional case remains to be investigated because in \cite{4} we have been able to compute the force only approximately, under the assumption that the charges are close to each other.  

The field $\varphi$ is the sum  of $\varphi_{cloud}$,  given by the exact  integral formula  (19), and of the two logarithmic Coulomb terms.   In the one dimensional case the pertinent field is known analytically \cite{4}, while in three dimensions we only have an approximate formula for it.  Note that formula (19) provides quite convenient starting point for numerical computations of  the field. 

2. The total screening of the charges  coupled to the signum-Gordon scalar field may resemble the phenomenon of total screening of external color charges interacting with a classical Yang-Mills field \cite{1}, \cite{2}. One should however note that there are several differences: 
in the Yang-Mills case the total screening  is proven for  external charges that are spatially extended; the screening field is time-dependent; the complete screening appears only  in a certain limit. Moreover,  the analytic form of the screening field is not known.  It is not clear to us whether the two cases of the total screening are  interrelated in some way. 

3.   In paper \cite{10} it is shown that a  classical scalar  field in the presence of point-like external charges can be utilized in order to unravel  certain essential features of the corresponding quantum theory, like asymptotic freedom or triviality, depending on the sign of coupling constant.  The main role is played by a classical perturbative solution that has the form of a formal series in powers of the coupling constant. The  field studied there  is the real, massless scalar field with the self-interaction  of the form $\lambda \varphi^4$.     Similar investigations in the cases of a  massive scalar field and Yang-Mills field are presented in \cite{11}, \cite{12}, respectively.  We think it would be interesting  to employ such a method to the signum-Gordon model, in particular because very little is known about  the properties of the quantum version of this model.

\end{document}